\documentclass[prd,preprintnumbers,twocolumn,amsmath,nofootinbib,amssymb]{revtex4}
\usepackage{graphicx,color,dcolumn,booktabs,bm}
\usepackage{longtable,lscape}
\usepackage{txfonts}
\usepackage{overpic}
\usepackage{amssymb}
\usepackage{epstopdf}
\usepackage{indentfirst}
\usepackage{feynmf}   
\usepackage{slashed}  
\usepackage{cases}
\usepackage{color}
\usepackage{float}
\usepackage{multirow}
\usepackage{ulem}
\usepackage{graphicx,color,dcolumn,booktabs,bm}
\usepackage{epsfig,dsfont,amssymb,amsmath,amsfonts,amsbsy,mathrsfs}

\graphicspath{{Figures/}} %

\usepackage{hyperref}
\hypersetup{colorlinks,citecolor=blue,anchorcolor=red,menucolor=red, linkcolor=red,filecolor=red,runcolor=red,urlcolor=blue,frenchlinks=red}

\makeatletter
\@addtoreset{equation}{section}
\makeatother

\allowdisplaybreaks

\begin{document}

\title{Hidden-charm pentaquarks with triple strangeness due to the $\Omega_{c}^{(*)}\bar{D}_s^{(*)}$ interactions}

\author{Fu-Lai Wang$^{1,2}$}
\email{wangfl2016@lzu.edu.cn}
\author{Xin-Dian Yang$^{1,2}$}
\email{yangxd20@lzu.edu.cn}
\author{Rui Chen$^{4,5}$}
\email{chen rui@pku.edu.cn}
\author{Xiang Liu$^{1,2,3}$\footnote{Corresponding author}}
\email{xiangliu@lzu.edu.cn}
\affiliation{$^1$School of Physical Science and Technology, Lanzhou University, Lanzhou 730000, China\\
$^2$Research Center for Hadron and CSR Physics, Lanzhou University and Institute of Modern Physics of CAS, Lanzhou 730000, China\\
$^3$Lanzhou Center for Theoretical Physics, Key Laboratory of Theoretical Physics of Gansu Province, and Frontiers Science Center for Rare Isotopes, Lanzhou University, Lanzhou 730000, China\\
$^4$Center of High Energy Physics, Peking University, Beijing 100871, China\\
$^5$School of Physics and State Key Laboratory of Nuclear Physics and Technology, Peking University, Beijing 100871, China}

\begin{abstract}
Motivated by the successful interpretation of these observed $P_c$ and $P_{cs}$ states under the meson-baryon molecular picture, we systematically investigate the possible hidden-charm molecular pentaquark states with triple strangeness which is due to the $\Omega_{c}^{(*)}\bar{D}_s^{(*)}$ interactions. We perform a dynamical calculation of the possible hidden-charm molecular pentaquarks with triple strangeness by the one-boson-exchange model, where the $S$-$D$ wave mixing effect and the coupled channel effect are taken into account in our calculation. Our results suggest that the $\Omega_{c}\bar D_s^*$ state with $J^P={3}/{2}^{-}$ and the $\Omega_{c}^{*}\bar D_s^*$ state with $J^P={5}/{2}^{-}$ can be recommended as the candidates of the hidden-charm molecular pentaquark with triple strangeness. Furthermore, we discuss the two-body hidden-charm strong decay behaviors of these possible hidden-charm molecular pentaquarks with triple strangeness by adopting the quark-interchange model. These predictions are expected to be tested at the LHCb, which can be as a potential research issue with more accumulated experimental data in near future.
\end{abstract}

\maketitle

\section{Introduction}\label{sec1}
As is well known, the study of the matter spectrum is an important way to explore the relevant matter structures and the involved interaction properties. In the hadron physics, since the discovery of the $X(3872)$ by the Belle Collaboration \cite{Choi:2003ue}, a series of exotic states has been observed benefiting from the accumulation of more and more experimental data with high precision, and the exotic hadrons have stimulated extensive studies in the past two decades (see the review articles \cite{Chen:2016qju,Liu:2019zoy,Olsen:2017bmm,Guo:2017jvc,Liu:2013waa,Hosaka:2016pey,Brambilla:2019esw} for learning the relevant processes). Exploring these exotic hadronic states not only gives new insights for revealing the hadron structures, but also provides useful hints to deepening our understanding of the nonperturbative behavior of the quantum chromodynamics (QCD) in the low energy regions.

In fact, investigating the pentaquark states has been a long history, which can be tracked back to the birth of the quark model \cite{GellMann:1964nj,Zweig:1981pd}. Among exotic hadronic states, the hidden-charm molecular pentaquarks have attracted much attention as early as 2010 \cite{Li:2014gra,Karliner:2015ina,Wu:2010jy,Wang:2011rga,Yang:2011wz,Wu:2012md,Chen:2015loa} and become a hot topic with the discovery of the $P_c(4380)$ and $P_c(4450)$ in the $\Lambda_b \to J/\psi p K$ process by the LHCb Collaboration \cite{Aaij:2015tga}. In 2019, there was a new progress about the observation of three narrow structures [$P_c(4312)$, $P_c(4440)$, and $P_c(4457)$] by revisiting the process $\Lambda_b \to J/\psi p K$ based on more collected data \cite{Aaij:2019vzc}, and they are just below the corresponding thresholds of the $S$-wave charmed baryon and $S$-wave anticharmed meson. This provides strong evidence to support the existence of the hidden-charm meson-baryon molecular states. More recently, the LHCb Collaboration reported a possible hidden-charm pentaquark with strangeness $P_{cs}(4459)$ \cite{Aaij:2020gdg}, and this structure can be assigned as the $\Xi_c \bar D^*$ molecular state \cite{Chen:2016ryt,Wu:2010vk,Hofmann:2005sw,Anisovich:2015zqa,Wang:2015wsa,Feijoo:2015kts,Lu:2016roh,Xiao:2019gjd,Shen:2020gpw,Chen:2015sxa,Zhang:2020cdi,Wang:2019nvm,Chen:2020uif,Peng:2020hql,1830432,1830426,Liu:2020hcv,1839195}.

Facing the present status of exploring the hidden-charm molecular pentaquarks \cite{Chen:2016qju,Liu:2019zoy,Olsen:2017bmm,Guo:2017jvc}, we naturally propose a meaningful question: why are we interested in the hidden-charm molecular pentaquark states? The hidden-charm pentaquark states are relatively easy to produce via the bottom baryon weak decays in the experimental facilities \cite{Aaij:2019vzc,Aaij:2020gdg}, and the hidden-charm quantum number is a crucial condition for the existence of the hadronic molecules \cite{Li:2014gra,Karliner:2015ina}. In addition, it is worth indicating that the heavy hadrons are more likely to generate the bound states due to the relatively small kinetic terms, and the interactions between the charmed baryon and the anticharmed meson may be mediated by exchanging a series of allowed light mesons \cite{Chen:2016qju,Liu:2019zoy}. Indeed, these announced hidden-charm pentaquark states have a ($c \bar c$) pair \cite{Chen:2016qju,Liu:2019zoy,Olsen:2017bmm,Guo:2017jvc}.

Based on the present research progress on the hidden-charm pentaquarks \cite{Chen:2016qju,Liu:2019zoy,Olsen:2017bmm,Guo:2017jvc}, the theorists should pay more attention to making the reliable prediction of various types of the hidden-charm molecular pentaquarks and give more abundant suggestions to searching for the hidden-charm molecular pentaquarks accessible at the forthcoming experiment. Generally speaking, there are two important approaches to construct the family of the hidden-charm molecular pentaquark states which is very special in the hadron spectroscopy. Firstly, we propose that there may exist a series of hidden-charm molecular pentaquarks with the different strangeness. Secondly, we also have enough reason to believe that there may exist more hidden-charm molecular pentaquark states with higher mass. In fact, we already studied the $\Xi_c^{(\prime,*)} \bar D_s^{(*)}$ systems with double strangeness \cite{Wang:2020bjt} and the $\mathcal{B}_{c}^{(*)} \bar T$ systems with $\mathcal{B}_{c}^{(*)}=\Lambda_c/\Sigma_c^{(*)}$ and $\bar T=\bar D_1/\bar D_2^*$ \cite{Wang:2019nwt}, and predicted a series of possible candidates of the hidden-charm molecular pentaquarks.
{In fact, the triple-strangeness hidden-charm pentaquarks may be regarded as systems that can be used to reveal the binding mechanism and the importance of the scalar-meson exchange as they are not expected to exist in the treatment of Ref. \cite{1839195}.}
Thus, we investigate the possible hidden-charm molecular pentaquarks with triple strangeness from the $\Omega_{c}^{(*)}\bar{D}_s^{(*)}$ interactions, which will be a main task of the present work.

In the present work, we perform a dynamical calculation with the possible hidden-charm molecular pentaquark states with triple strangeness by adopting the one-boson-exchange (OBE) model \cite{Chen:2016qju,Liu:2019zoy}, which involves the interactions between an $S$-wave charmed baryon $\Omega_{c}^{(*)}$ and an $S$-wave anticharmed-strange meson $\bar{D}_s^{(*)}$. In concrete calculation, the $S$-$D$ wave mixing effect and the coupled channel effect are taken into account. Furthermore, we study the two-body hidden-charm strong decay behaviors of these possible hidden-charm molecular pentaquarks. Here, we adopt the quark-interchange model to estimate the transition amplitudes for the decay widths \cite{Barnes:1991em, Barnes:1999hs, Barnes:2000hu}, which is widely used to give the decay information of the exotic hadronic states during the last few decades \cite{Wang:2018pwi,Wang:2019spc,Xiao:2019spy,Wang:2020prk,Hilbert:2007hc}. We hope that the present investigation is a key step to complement the family of the hidden-charm molecular pentaquark state and may provide crucial information of searching for possible hidden-charm molecular pentaquarks with triple strangeness. With higher statistic data accumulation at Run III of the LHC and after High-Luminosity-LHC upgrade \cite{Bediaga:2018lhg}, it is highly probable that these possible hidden-charm molecular pentaquarks with triple strangeness can be detected at the LHCb Collaboration in the near future, which will be full of opportunities and challenges.

The remainder of this paper is organized as follows. In Sec. \ref{sec2}, we introduce how to deduce the effective potentials and present the bound state properties of these investigated $\Omega_{c}^{(*)} \bar{D}_s^{(*)}$ systems. We present the  quark-interchange model and the two-body strong decay behaviors of these possible molecular pentaquarks in Sec. \ref{sec3}. Finally, a short summary follows in Sec. \ref{sec4}.

\section{The $\Omega_{c}^{(*)}\bar{D}_s^{(*)}$ interactions}\label{sec2}
\subsection{OBE effective potentials}
In the present work, we study the interactions between an $S$-wave charmed baryon $\Omega_{c}^{(*)}$ and an $S$-wave anticharmed-strange meson $\bar{D}_s^{(*)}$. Here, we adopt the OBE model \cite{Chen:2016qju,Liu:2019zoy}, and consider the effective potentials from the $f_0(980)$, $\eta$, and $\phi$ exchanges.
{In particular, we need to emphasize that the light scalar meson $f_0(980)$ exchange provides effective interaction for these investigated systems, and we do not consider the $\sigma$ and $a_0(980)$ exchanges in our calculation, since the $\sigma$ is usually considered as a meson with only up and down quarks and the $a_0(980)$ is the light isovector scalar meson.}
In this subsection, we construct the relevant wave functions and effective Lagrangians, and deduce the OBE effective potentials in the coordinate space for all of the investigated systems.

Firstly, we introduce the flavor and spin-orbital wave functions involved in our calculation. For the $\Omega_{c}^{(*)}\bar{D}_s^{(*)}$ systems, the flavor wave function $|I,I_3\rangle$ is quite simple and reads as $|0,0\rangle=|\Omega_{c}^{(*)0}{D}_s^{(*)-}\rangle$, where $I$ and $I_3$ are the isospin and its third component of the discussed system. In addition, the spin-orbital wave functions $|{}^{2S+1}L_{J}\rangle$ for these investigated $\Omega_{c}^{(*)}\bar{D}_s^{(*)}$ systems are explicitly written as
\begin{eqnarray}
\left|\Omega_{c}\bar{D}_{s}\left({}^{2S+1}L_{J}\right)\right\rangle&=&\sum_{m,m_L}C^{J,M}_{\frac{1}{2}m,Lm_L}\chi_{\frac{1}{2}m}\left|Y_{L,m_L}\right\rangle,\nonumber\\
\left|\Omega_{c}^*\bar{D}_{s}\left({}^{2S+1}L_{J}\right)\right\rangle&=&\sum_{m,m_L}C^{J,M}_{\frac{3}{2}m,Lm_L}\Phi_{\frac{3}{2}m}\left|Y_{L,m_L}\right\rangle,\nonumber\\
\left|\Omega_{c}\bar{D}_{s}^*\left({}^{2S+1}L_{J}\right)\right\rangle&=&\sum_{m,m',m_S,m_L}C^{S,m_S}_{\frac{1}{2}m,1m'}C^{J,M}_{Sm_S,Lm_L}\chi_{\frac{1}{2}m}\epsilon_{m'}^{\mu}\left|Y_{L,m_L}\right\rangle,\nonumber\\
\left|\Omega_{c}^{*}\bar{D}_{s}^*\left({}^{2S+1}L_{J}\right)\right\rangle&=&\sum_{m,m',m_S,m_L}C^{S,m_S}_{\frac{3}{2}m,1m'}C^{J,M}_{Sm_S,Lm_L}\Phi_{\frac{3}{2}m}\epsilon_{m'}^{\mu}\left|Y_{L,m_L}\right\rangle.\nonumber\\
\end{eqnarray}
In the above expressions, $S$, $L$, and $J$ denote the spin, orbit angular momentum, and total angular momentum for the discussed system, respectively. The constant $C^{e,f}_{ab,cd}$ is the Clebsch-Gordan coefficient, and $|Y_{L,m_L}\rangle$ is the spherical harmonics function. In the static limit, the polarization vector $\epsilon_{m}^{\mu}\,(m=0,\,\pm1)$ with the spin-1 field can be expressed as $\epsilon_{0}^{\mu}= \left(0,0,0,-1\right)$ and $\epsilon_{\pm}^{\mu}= \left(0,\,\pm1,\,i,\,0\right)/\sqrt{2}$. $\chi_{\frac{1}{2}m}$ stands for the spin wave function of the charmed baryon with spin $S={1}/{2}$, and the polarization tensor $\Phi_{\frac{3}{2}m}$ of the charmed baryon with spin quantum number $S={3}/{2}$ can be written in a general form, i.e.,
\begin{eqnarray}
\Phi_{\frac{3}{2}m}=\sum_{m_1,m_2}C^{\frac{3}{2},m}_{\frac{1}{2}m_1,1m_2}\chi_{\frac{1}{2}m_1}\epsilon_{m_2}^{\mu}.
\end{eqnarray}

In order to write out the relevant scattering amplitudes quantitatively, we usually adopt the effective Lagrangian approach. To be convenient, we construct two types of super-fields $\mathcal{S}_{\mu}$ and $H^{(\overline{Q})}_a$ via the heavy quark limit \cite{Wise:1992hn}. The superfield $\mathcal{S}_{\mu}$ is expressed as a combination of the charmed baryons $\mathcal{B}_6$ with $J^P=1/2^+$ and $\mathcal{B}^*_6$ with $J^P=3/2^+$ in the $6_F$ flavor representation \cite{Chen:2017xat}, and the superfield $H^{(\overline{Q})}_a$ includes the anticharmed-strange  vector meson $\bar{D}^{*}_s$ with $J^P=1^-$ and the pseudoscalar meson $\bar{D}_s$ with $J^P=0^-$ \cite{Ding:2008gr}. The general expressions of the super-fields $\mathcal{S}_{\mu}$ and $H^{(\overline{Q})}_a$ can be given by
\begin{eqnarray}
\mathcal{S}_{\mu}&=&-\sqrt{\frac{1}{3}}(\gamma_{\mu}+v_{\mu})\gamma^5\mathcal{B}_6+\mathcal{B}_{6\mu}^*,\nonumber\\
H^{(\overline{Q})}_a&=&\left(\bar{D}^{*(\overline{Q})\mu}_{a}\gamma_{\mu}-\bar{D}^{(\overline{Q})}_a\gamma_5\right)\frac{1-\slash \!\!\!v}{2}.
\end{eqnarray}
Here, $v_{\mu}=(1,\bm{0})$ is the four velocity under the nonrelativistic approximation.

With the above preparation, we construct the relevant effective Lagrangians to describe the interactions among the heavy hadrons $\mathcal{B}_6^{(*)}/\bar{D}_s^{(*)}$ and the light scalar, pseudoscalar, or vector mesons
as \cite{Ding:2008gr,Chen:2017xat}
\begin{eqnarray}
\mathcal{L}_{\mathcal{B}^{(*)}_6} &=&  l_S\langle\bar{\mathcal{S}}_{\mu}f_0\mathcal{S}^{\mu}\rangle
         -\frac{3}{2}g_1\varepsilon^{\mu\nu\lambda\kappa}v_{\kappa}\langle\bar{\mathcal{S}}_{\mu}{\mathcal A}_{\nu}\mathcal{S}_{\lambda}\rangle\nonumber\\
  &&+i\beta_{S}\langle\bar{\mathcal{S}}_{\mu}v_{\alpha}\left(\mathcal{V}^{\alpha}-\rho^{\alpha}\right) \mathcal{S}^{\mu}\rangle
         +\lambda_S\langle\bar{\mathcal{S}}_{\mu}F^{\mu\nu}(\rho)\mathcal{S}_{\nu}\rangle,\nonumber\\
\mathcal{L}_{H}&=&g_S\langle \bar{H}^{(\overline{Q})}_a f_0 H^{(\overline{Q})}_a\rangle+ig\langle \bar{H}^{(\overline{Q})}_a\gamma_{\mu}{\mathcal A}_{ab}^{\mu}\gamma_5H^{(\overline{Q})}_b\rangle\nonumber\\
  &&-i\beta\langle \bar{H}^{(\overline{Q})}_av_{\mu}\left(\mathcal{V}^{\mu}-\rho^{\mu}\right)_{ab}H^{(\overline{Q})}_b\rangle\nonumber\\
  &&+i\lambda\langle \bar{H}^{(\overline{Q})}_a\sigma_{\mu\nu}F^{\mu\nu}(\rho)_{ab}H^{(\overline{Q})}_b\rangle,
\end{eqnarray}
which satisfy the requirement of the heavy quark symmetry, the chiral symmetry, and the hidden local symmetry \cite{Casalbuoni:1992gi,Casalbuoni:1996pg,Yan:1992gz,Harada:2003jx,Bando:1987br}. The axial current $\mathcal{A}_\mu$ and the vector current ${\cal V}_{\mu}$ can be defined as ${\mathcal A}_{\mu}=\left(\xi^{\dagger}\partial_{\mu}\xi-\xi\partial_{\mu}\xi^{\dagger}\right)/2$ and ${\mathcal V}_{\mu}=\left(\xi^{\dagger}\partial_{\mu}\xi+\xi\partial_{\mu}\xi^{\dagger}\right)/2$, respectively. Here, the pseudo-Goldstone field can be written as $\xi=\exp(i\mathbb{P}/f_\pi)$, where $f_\pi$ is the pion decay constant. In the above formulas, the vector meson field $\rho_{\mu}$ and its strength tensor $F_{\mu\nu}(\rho)$ are $\rho_{\mu}=i{g_V}\mathbb{V}_{\mu}/{\sqrt{2}}$ and $F_{\mu\nu}(\rho)=\partial_{\mu}\rho_{\nu}-\partial_{\nu}\rho_{\mu}+[\rho_{\mu},\rho_{\nu}]$, respectively. Here, $\mathcal{B}_6^{(*)}$, $\mathbb{V}_{\mu}$, and ${\mathbb{P}}$ are the matrices of the charmed baryon in the $6_F$ flavor representation, light vector meson, and light pseudoscalar meson, which can be written as
\begin{eqnarray}
\left.\begin{array}{c}
\mathcal{B}_6^{(*)} = \left(\begin{array}{ccc}
         \Sigma_c^{{(*)}++}                  &\frac{\Sigma_c^{{(*)}+}}{\sqrt{2}}     &\frac{\Xi_c^{(',*)+}}{\sqrt{2}}\\
         \frac{\Sigma_c^{{(*)}+}}{\sqrt{2}}      &\Sigma_c^{{(*)}0}    &\frac{\Xi_c^{(',*)0}}{\sqrt{2}}\\
         \frac{\Xi_c^{(',*)+}}{\sqrt{2}}    &\frac{\Xi_c^{(',*)0}}{\sqrt{2}}      &\Omega_c^{(*)0}
\end{array}\right),\\
{\mathbb{V}}_{\mu} = {\left(\begin{array}{ccc}
       \frac{\rho^0}{\sqrt{2}}+\frac{\omega}{\sqrt{2}} &\rho^+ &K^{*+}\\
       \rho^-       &-\frac{\rho^0}{\sqrt{2}}+\frac{\omega}{\sqrt{2}} &K^{*0}\\
       K^{*-}         &\bar K^{*0}   & \phi     \end{array}\right)}_{\mu},\\
{\mathbb{P}} = {\left(\begin{array}{ccc}
       \frac{\pi^0}{\sqrt{2}}+\frac{\eta}{\sqrt{6}} &\pi^+ &K^+\\
       \pi^-       &-\frac{\pi^0}{\sqrt{2}}+\frac{\eta}{\sqrt{6}} &K^0\\
       K^-         &\bar K^0   &-\sqrt{\frac{2}{3}} \eta     \end{array}\right)},
\end{array}\right.
\end{eqnarray}
respectively.
By expanding the compact effective Lagrangians to the leading order of the pseudo-Goldstone field $\xi$, we can further obtain the concrete effective Lagrangians. The effective Lagrangians for $\mathcal{B}_6^{(*)}$ and the light mesons are expressed as
\begin{eqnarray}
\mathcal{L}_{\mathcal{B}_{6}^{(*)}\mathcal{B}_{6}^{(*)}f_0} &=&-l_S\langle\bar{\mathcal{B}}_6 f_0\mathcal{B}_6\rangle+l_S\langle\bar{\mathcal{B}}_{6\mu}^{*}f_0\mathcal{B}_6^{*\mu}\rangle\nonumber\\
    &&-\frac{l_S}{\sqrt{3}}\langle\bar{\mathcal{B}}_{6\mu}^{*}f_0 \left(\gamma^{\mu}+v^{\mu}\right)\gamma^5\mathcal{B}_6\rangle+h.c.,\label{n1}\\
\mathcal{L}_{\mathcal{B}_6^{(*)}\mathcal{B}_6^{(*)}\mathbb{P}} &=&i\frac{g_1}{2f_{\pi}}\varepsilon^{\mu\nu\lambda\kappa}v_{\kappa}\langle\bar{\mathcal{B}}_6 \gamma_{\mu}\gamma_{\lambda}\partial_{\nu}\mathbb{P}\mathcal{B}_6\rangle\nonumber\\
    &&-i\frac{3g_1}{2f_{\pi}}\varepsilon^{\mu\nu\lambda\kappa}v_{\kappa}\langle\bar{\mathcal{B}}_{6\mu}^{*}\partial_{\nu}\mathbb{P}\mathcal{B}_{6\lambda}^*\rangle\nonumber\\
    &&+i\frac{\sqrt{3}g_1}{2f_{\pi}}v_{\kappa}\varepsilon^{\mu\nu\lambda\kappa}\langle\bar{\mathcal{B}}_{6\mu}^*\partial_{\nu}\mathbb{P}{\gamma_{\lambda}\gamma^5}\mathcal{B}_6\rangle+h.c.,\\
\mathcal{L}_{\mathcal{B}_6^{(*)}\mathcal{B}_6^{(*)}\mathbb{V}}&=&-\frac{\beta_Sg_V}{\sqrt{2}}\langle\bar{\mathcal{B}}_6v\cdot\mathbb{V}\mathcal{B}_6\rangle\nonumber\\
    &&-i\frac{\lambda_S g_V}{3\sqrt{2}}\langle\bar{\mathcal{B}}_6\gamma_{\mu}\gamma_{\nu}\left(\partial^{\mu}\mathbb{V}^{\nu}-\partial^{\nu}\mathbb{V}^{\mu}\right)\mathcal{B}_6\rangle\nonumber\\
    &&-\frac{\beta_Sg_V}{\sqrt{6}}\langle\bar{\mathcal{B}}_{6\mu}^*v\cdot \mathbb{V}\left(\gamma^{\mu}+v^{\mu}\right)\gamma^5\mathcal{B}_6\rangle\nonumber\\
    &&-i\frac{\lambda_Sg_V}{\sqrt{6}}\langle\bar{\mathcal{B}}_{6\mu}^*\left(\partial^{\mu}\mathbb{V}^{\nu}-\partial^{\nu}\mathbb{V}^{\mu}\right)\left(\gamma_{\nu}+v_{\nu}\right)\gamma^5\mathcal{B}_6\rangle\nonumber\\
    &&+\frac{\beta_Sg_V}{\sqrt{2}}\langle\bar{\mathcal{B}}_{6\mu}^*v\cdot {V}\mathcal{B}_6^{*\mu}\rangle\nonumber\\
    &&+i\frac{\lambda_Sg_V}{\sqrt{2}}\langle\bar{\mathcal{B}}_{6\mu}^*\left(\partial^{\mu}\mathbb{V}^{\nu}-\partial^{\nu}\mathbb{V}^{\mu}\right)\mathcal{B}_{6\nu}^*\rangle+h.c.,
\end{eqnarray}
and the effective Lagrangians to describe the $S$-wave anticharmed-strange mesons $\bar{D}_s^{(*)}$ and the light scalar, pseudoscalar, or vector mesons are
\begin{eqnarray}
\mathcal{L}_{{\bar D}^{(*)}{\bar D}^{(*)}f_0} &=&-2g_S{\bar D}_a {\bar D}_a^{\dag} f_0+ 2g_S {\bar D}_{a\mu}^* {\bar D}_a^{*\mu\dag} f_0,\label{n2}\\
\mathcal{L}_{{\bar D}^{(*)}{\bar D}^{(*)}\mathbb{P}}&=&\frac{2ig}{f_{\pi}}v^{\alpha}\varepsilon_{\alpha\mu\nu\lambda}{\bar D}_a^{*\mu\dag}{\bar D}_b^{*\lambda}\partial^{\nu}{\mathbb{P}}_{ab}\nonumber\\
    &&+\frac{2g}{f_{\pi}}\left({\bar D}_a^{*\mu\dag}{\bar D}_b+{\bar D}_a^{\dag}{\bar D}_b^{*\mu}\right)\partial_{\mu}{\mathbb{P}}_{ab},\\
\mathcal{L}_{{\bar D}^{(*)}{\bar D}^{(*)}\mathbb{V}} &=&\sqrt{2}\beta g_V {\bar D}_a {\bar D}_b^{\dag} v\cdot\mathbb{V}_{ab}-\sqrt{2}\beta g_V {\bar D}_{a\mu}^* {\bar D}_b^{*\mu\dag}v\cdot\mathbb{V}_{ab}\nonumber\\
    &&-2\sqrt{2}i\lambda g_V {\bar D}_a^{*\mu\dag}{\bar D}_b^{*\nu}\left(\partial_{\mu}\mathbb{V}_{\nu}-\partial_{\nu}\mathbb{V}_{\mu}\right)_{ab}\nonumber\\
    &&-2\sqrt{2}\lambda g_V v^{\lambda}\varepsilon_{\lambda\mu\alpha\beta}\left({\bar D}_a^{*\mu\dag}{\bar D}_b+{\bar D}_a^{\dag}{\bar D}_b^{*\mu}\right)\partial^{\alpha}\mathbb{V}^{\beta}_{ab}.\nonumber\\
\end{eqnarray}

In the above effective Lagrangians, the coupling constants can be either extracted from the experimental data or calculated by the theoretical models, and the signs of these coupling constants are fixed via the quark model \cite{Riska:2000gd}. The values of these coupling constants are $l_S=6.20$, $g_S=0.76$,\footnote{In this work, we consider the contribution from light scalar meson $f_0(980)$ exchange. Here, the corresponding coupling constant involved in effective Lagrangians [Eq. (\ref{n1}) and Eq. (\ref{n2})] is approximately taken as the same as that for the case of light scalar $\sigma$.} $g_1=0.94$, $g=0.59$, $f_\pi=132~\rm{MeV}$, $\beta_S g_V=10.14$, $\beta g_V=-5.25$, $\lambda_S g_V=19.2~\rm{GeV}^{-1}$, and $\lambda g_V =-3.27~\rm{GeV}^{-1}$ \cite{Chen:2019asm}, which are widely used to discuss the hadronic molecular states \cite{Wang:2020bjt,Chen:2017xat,Wang:2019nwt,Chen:2019asm,He:2015cea,He:2019ify,Chen:2018pzd}. In particular, we need to emphasize that these input coupling constants can well reproduce the masses of the $P_c(4312)$, $P_c(4440)$, and $P_c(4457)$ \cite{Aaij:2019vzc} under the meson-baryon molecular picture when adopting the OBE model \cite{Chen:2019asm,He:2019ify}.

\begin{figure}[!htbp]
\includegraphics[width=4.0cm,keepaspectratio]{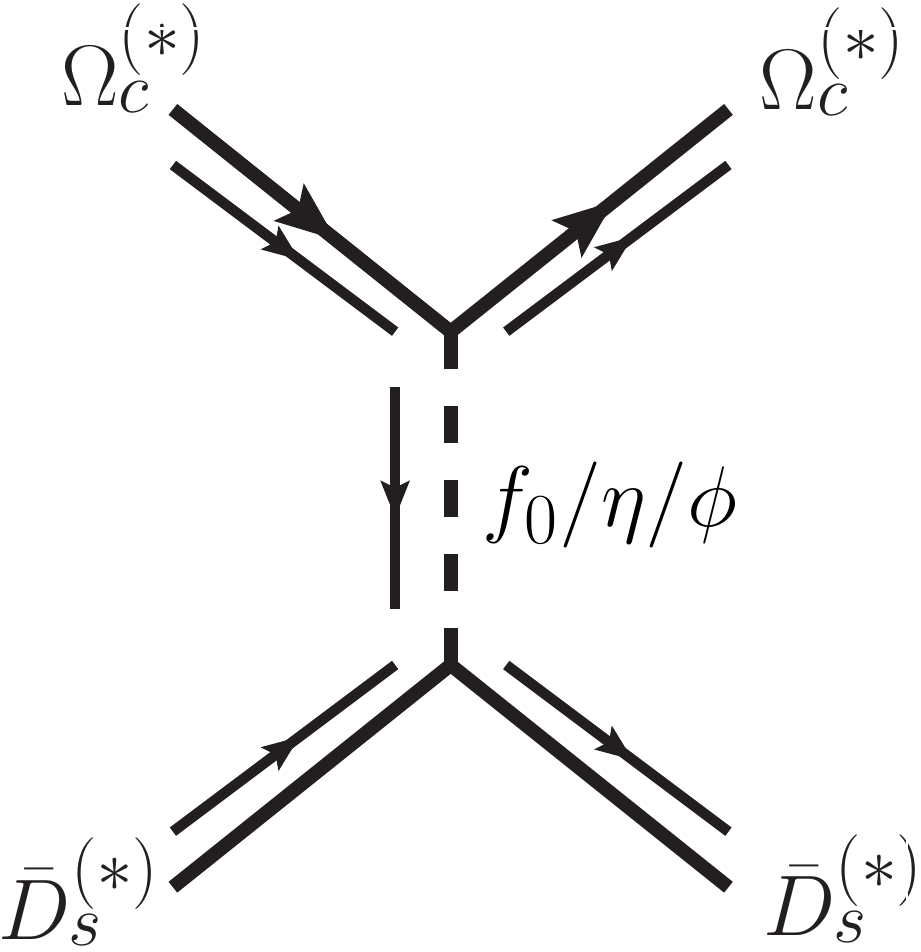}
\caption{Relevant Feynman diagram for the $\Omega_{c}^{(*)}\bar {D}_s^{(*)} \to \Omega_{c}^{(*)}\bar {D}_s^{(*)}$ scattering processes.}
\label{Feynmandiagram}
\end{figure}

We follow the standard strategy to deduce the effective potentials in the coordinate space in Refs. \cite{Wang:2020dya,Wang:2019nwt,Wang:2019aoc}, which is a lengthy and tedious calculation. In Fig. \ref{Feynmandiagram}, we present the relevant Feynman diagram for the $\Omega_{c}^{(*)}\bar {D}_s^{(*)} \to \Omega_{c}^{(*)}\bar {D}_s^{(*)}$ scattering processes.

At the hadronic level, we firstly write out the scattering amplitude $\mathcal{M}(h_1h_2\to h_3h_4)$ of the scattering process $h_1h_2\to h_3h_4$ by considering the effective Lagrangian approach. And then, the effective potential in momentum space $\mathcal{V}^{h_1h_2\to h_3h_4}(\bm{q})$ can be related to the scattering amplitude $\mathcal{M}(h_1h_2\to h_3h_4)$ with the help of the Breit approximation \cite{Breit:1929zz,Breit:1930zza} and the nonrelativistic normalization, i.e.,
\begin{eqnarray}\label{breit}
\mathcal{V}_E^{h_1h_2\to h_3h_4}(\bm{q}) &=&-\frac{\mathcal{M}(h_1h_2\to h_3h_4)} {\sqrt{\prod_i2m_i\prod_f2m_f}},
\end{eqnarray}
where $m_i$ and $m_f$ are the masses of the initial states $(h_1, \,h_2)$ and final states $(h_3, \,h_4)$, respectively. By performing the Fourier transformation, the effective potential in the coordinate space $\mathcal{V}^{h_1h_2\to h_3h_4}(\bm{r})$ can be deduced
\begin{eqnarray}
\mathcal{V}^{h_1h_2\to h_3h_4}_E(\bm{r}) =\int\frac{d^3\bm{q}}{(2\pi)^3}e^{i\bm{q}\cdot\bm{r}}\mathcal{V}_E^{h_1h_2\to h_3h_4}(\bm{q})\mathcal{F}^2(q^2,m_E^2).\nonumber\\
\end{eqnarray}
In order to reflect the finite size effect of the discussed hadrons and compensate the off-shell effect of the exchanged light mesons \cite{Wang:2020dya}, we need to introduce the monopole form factor $\mathcal{F}(q^2,m_E^2) = (\Lambda^2-m_E^2)/(\Lambda^2-q^2)$ in the interaction vertex \cite{Tornqvist:1993ng,Tornqvist:1993vu}. Here, $\Lambda$, $m_E$, and $q$ are the cutoff parameter, the mass, and the four momentum of the exchanged light meson, respectively.

In addition, we also need a series of normalization relations for the heavy hadrons $D_{s}$, $D_{s}^{*}$, $\Omega_{c}$, and $\Omega_{c}^{*}$, i.e.,
\begin{eqnarray}
\langle 0|D_{s}|c\bar{s}\left(0^-\right)\rangle&=&\sqrt{M_{D_{s}}},\\
\langle 0|D_{s}^{*\mu}|c\bar{s}\left(1^-\right)\rangle&=&\sqrt{M_{D_{s}^*}}\epsilon^\mu,\\
\langle 0|\Omega_{c}|css\left({1}/{2}^+\right)\rangle &=& \sqrt{2M_{\Omega_{c}}}{\left(\chi_{\frac{1}{2}m},\frac{\bf{\sigma}\cdot\bf{p}}{2M_{\Omega_{c}}}\chi_{\frac{1}{2}m}\right)^T},\\
\langle 0|\Omega_{c}^{*\mu}|css\left({3}/{2}^+\right)\rangle &=&\sum_{m_1,m_2}C_{1/2,m_1;1,m_2}^{3/2,m_1+m_2}\sqrt{2M_{\Omega_{c}^*}}\nonumber\\
  &&\times\left(\chi_{\frac{1}{2}m_1},\frac{\bf{\sigma}\cdot\bf{p}}{2M_{\Omega_{c}^*}}\chi_{\frac{1}{2}m_1}\right)^T\epsilon^{\mu}_{m_2}.
\end{eqnarray}

With the above preparation, we can deduce the OBE effective potentials in the coordinate space for all of the investigated processes, which are collected in the~\ref{app01}.

\subsection{Finding bound state solutions for discussed systems}
Now, we attempt to find the loosely bound state solutions of these discussed $\Omega_{c}^{(*)}\bar D_s^{(*)}$ systems by solving the coupled channel Schr\"{o}dinger equation, i.e.,
\begin{eqnarray}\label{SE}
-\frac{1}{2\mu}\left(\nabla^2-\frac{\ell(\ell+1)}{r^2}\right)\psi(r)+V(r)\psi(r)=E\psi(r)
\end{eqnarray}
with $\nabla^2=\frac{1}{r^2}\frac{\partial}{\partial r}r^2\frac{\partial}{\partial r}$, where $\mu=\frac{m_1m_2}{m_1+m_2}$ is the reduced mass for the discussed system. The bound state solutions include the binding energy $E$, the root-mean-square radius $r_{\rm RMS}$, and the probability of the individual channel $P_i$, which provides us with valuable information to analyze whether the loosely bound state exists. In this work, we are interested in the $S$-wave $\Omega_{c}^{(*)}\bar D_s^{(*)}$ systems since there exists the repulsive centrifugal potential for the higher partial wave states $\ell \geqslant 1$.

In our calculation, the masses of these involved hadrons are $m_{f_0}=990.00$ MeV, $m_\eta=547.85$ MeV, $m_\phi=1019.46$ MeV, $m_{D_s}=1968.34$ MeV, $m_{D_s^*}=2112.20$ MeV, $m_{\Omega_{c}}=2695.20$ MeV, and $m_{\Omega_{c}^*}=2765.90$ MeV, which are taken from the Particle Data Group (PDG) \cite{Zyla:2020zbs}.  As the remaining phenomenological parameter, we take the cutoff value from 1.00 to 4.00 GeV. Usually, a loosely bound state with the cutoff parameter closed to 1.00 GeV can be suggested as the possible hadronic molecular candidate according to the experience of the deuteron \cite{Tornqvist:1993ng,Tornqvist:1993vu,Wang:2019nwt,Chen:2017jjn}. For an ideal hadronic molecular candidate, the reasonable binding energy should be at most tens of MeV, and the typical size should be larger than the size of all the included component hadrons \cite{Chen:2017xat}.

In addition, the $S$-$D$ wave mixing effect is considered in this work, which plays an important role to modify the bound state properties of the deuteron \cite{Wang:2019nwt}. The relevant channels $|{}^{2S+1}L_{J}\rangle$ are summarized in Table~\ref{spin-orbit wave functions}.
\renewcommand\tabcolsep{0.13cm}
\renewcommand{\arraystretch}{1.50}
\begin{table}[!htpb]
\centering
\caption{The relevant channels $|{}^{2S+1}L_{J}\rangle$ involved in our calculation. Here, ``$...$" means that the $S$-wave component for the corresponding channel does not exist.}\label{spin-orbit wave functions}
\begin{tabular}{c|cccc}\toprule[1pt]\toprule[1pt]
 $J^{P}$      &$\Omega_{c}\bar{D}_s$ &$\Omega_{c}^{*}\bar{D}_s$&$\Omega_{c}\bar{D}_s^{*}$&$\Omega_{c}^{*}\bar{D}^{*}_s$\\
 \cline{1-5}
${1}/{2}^{-}$ &$|{}^2\mathbb{S}_{1/2}\rangle$&$...$ &$|{}^2\mathbb{S}_{1/2}\rangle/|{}^4\mathbb{D}_{1/2}\rangle$&$|{}^2\mathbb{S}_{1/2}\rangle/|{}^{4,6}\mathbb{D}_{1/2}\rangle$\\
${3}/{2}^{-}$ &$...$ &$|{}^4\mathbb{S}_{3/2}\rangle/|{}^{4}\mathbb{D}_{3/2}\rangle$&$|{}^4\mathbb{S}_{3/2}\rangle/|{}^{2,4}\mathbb{D}_{3/2}\rangle$&$|{}^4\mathbb{S}_{3/2}\rangle/|{}^{2,4,6}\mathbb{D}_{3/2}\rangle$\\
${5}/{2}^{-}$ &$...$&$...$&$...$&$|{}^6\mathbb{S}_{5/2}\rangle/|{}^{2,4,6}\mathbb{D}_{5/2}\rangle$\\
\bottomrule[1pt]\bottomrule[1pt]
\end{tabular}
\end{table}

Before performing numerical calculation, we analyze the OBE effective potentials for these discussed $\Omega_{c}^{(*)}\bar D_s^{(*)}$ systems as below:
\begin{itemize}
  \item For the $\Omega_{c}\bar D_s$ and $\Omega_{c}^*\bar D_s$ systems, only the $f_0$ and $\phi$ exchange interactions are allowed. Meanwhile, the tensor force from the $S$-$D$ wave mixing effect disappears in the effective potentials, and thus the contribution of the $S$-$D$ wave mixing effect does not affect the bound state properties of the $\Omega_{c}\bar D_s$ and $\Omega_{c}^*\bar D_s$ systems.
  \item For the $\Omega_{c}\bar D_s^*$ and $\Omega_{c}^*\bar D_s^*$ systems, in addition to the $f_0$ and $\phi$ exchange interactions, the $\eta$ exchange interaction and the $S$-$D$ wave mixing effect need to be taken into account.
\end{itemize}

\subsubsection{The $\Omega_{c}\bar D_s$ and $\Omega_{c}^*\bar D_s$  systems}
For the $S$-wave $\Omega_{c}\bar D_s$ state with $J^P={1}/{2}^{-}$ and the $S$-wave $\Omega_{c}^*\bar D_s$ state with $J^P={3}/{2}^{-}$, we fail to find their bound state solutions by varying the cutoff parameter in the range of $1.00$-$4.00~{\rm GeV}$ with the single channel analysis. Nevertheless, we can further take into account the coupled channel effect. In the coupled channel analysis, the binding energy of the bound state is determined by the lowest mass threshold among various investigated channels \cite{Chen:2017xat}.

For the $S$-wave $\Omega_{c}\bar D_s$ state with $J^P={1}/{2}^{-}$, we consider the coupled channel effect from the $\Omega_{c}\bar D_s$, $\Omega_{c}\bar D_s^*$, and $\Omega_{c}^{*}\bar D_s^*$ channels. In Table~\ref{r1}, we present the obtained bound state solutions by performing the coupled channel analysis. When we set the cutoff parameter $\Lambda$ around 2.92 GeV, the loosely bound state solutions can be obtained, and the $\Omega_{c}\bar D_s$ channel is dominant with almost 90\% probabilities. Since the cutoff parameter $\Lambda$ is obviously different from 1.00 GeV \cite{Tornqvist:1993ng,Tornqvist:1993vu}, the $S$-wave $\Omega_{c}\bar D_s$ state with $J^P={1}/{2}^{-}$ is not priority for recommending the hadronic molecular candidate.
\renewcommand\tabcolsep{0.46cm}
\renewcommand{\arraystretch}{1.50}
\begin{table}[!htbp]
\caption{Bound state solutions of the $S$-wave $\Omega_{c}\bar D_s$ state with $J^P={1}/{2}^{-}$ by performing coupled channel analysis. Here, the cutoff parameter $\Lambda$, binding energy $E$, and root-mean-square radius $r_{RMS}$ are in units of $ \rm{GeV}$, $\rm {MeV}$, and $\rm {fm}$, respectively.}\label{r1}
\begin{tabular}{cccc}\toprule[1pt]\toprule[1pt]
$\Lambda$ &$E$  &$r_{\rm RMS}$ &P($\Omega_{c}\bar D_s/\Omega_{c}\bar D_s^*/\Omega_{c}^{*}\bar D_s^*$)\\
\cline{1-4}
2.92&$-3.71$ &1.26&\textbf{92.92}/4.77/2.31\\
2.93&$-12.65$ &0.64&\textbf{90.11}/6.66/3.23\\
\bottomrule[1pt]\bottomrule[1pt]
\end{tabular}
\end{table}

In Table~\ref{r2}, we list the bound state solutions of the $S$-wave $\Omega_{c}^*\bar D_s$ state with $J^P={3}/{2}^{-}$ with the coupled channel analysis. Our numerical results show that the bound state solutions can be obtained by choosing the cutoff parameter $\Lambda$ around 1.78 GeV or even larger, and the $\Omega_{c}\bar D_s^*$ system is the dominant channel with the probabilities over 80\%. However, we find the size ($r_{\rm RMS} \sim 0.33~ {\rm {fm}}$) of this bound state is too small,\footnote{
{We notice that the obtained values of $r_{RMS}$ are too small, which is due to the fact that this sysmtem is dominated by the $\Omega_{c}\bar D_s^*$ channel as shown in the last column of Table \ref{r2}.}}
which is not consistent with a loosely molecular state picture \cite{Chen:2017xat}. Thus, we tentatively exclude the possibility of the existence of the $S$-wave $\Omega_{c}^*\bar D_s$ molecular state with $J^P={3}/{2}^{-}$.
\renewcommand\tabcolsep{0.46cm}
\renewcommand{\arraystretch}{1.50}
\begin{table}[!htbp]
\caption{Bound state solutions of the $S$-wave $\Omega_{c}^*\bar D_s$ state with $J^P={3}/{2}^{-}$ when the coupled channel effect is introduced. The units are the same as Table~\ref{r1}.}\label{r2}
\begin{tabular}{cccc}\toprule[1pt]\toprule[1pt]
$\Lambda$ &$E$  &$r_{\rm RMS}$ &P($\Omega_{c}^*\bar D_s/\Omega_{c}\bar D_s^*/\Omega_{c}^{*}\bar D_s^*$)\\
\cline{1-4}
1.78&$-6.15$ &0.33&0.01/\textbf{86.64}/13.36\\
1.79&$-17.41$ &0.32&0.01/\textbf{86.37}/13.63\\
\bottomrule[1pt]\bottomrule[1pt]
\end{tabular}
\end{table}

\subsubsection{The $\Omega_{c}\bar D_s^*$ and $\Omega_{c}^{*}\bar D_s^*$ systems}
For the $S$-wave $\Omega_{c}\bar D_s^*$ system, the relevant numerical results are collected in Table~\ref{r3}. For $J^P={1}/{2}^{-}$, there do not exist bound states until
we increase the cutoff parameter to be around 4.00 GeV, even if we consider the coupled channel effect. Thus, we conclude that our quantitative analysis does not support the existence of the $S$-wave $\Omega_{c}\bar D_s^*$ molecular state with $J^P={1}/{2}^{-}$.
\renewcommand\tabcolsep{0.28cm}
\renewcommand{\arraystretch}{1.50}
\begin{table*}[!htbp]
\caption{Bound state solutions of the $S$-wave $\Omega_{c}\bar D_s^*$ system. The units are the same as Table~\ref{r1}.}\label{r3}
\begin{tabular}{c|ccc|cccc|cccc}\toprule[1.0pt]\toprule[1.0pt]
\multicolumn{1}{c|}{Effect}&\multicolumn{3}{c|}{Single channel}&\multicolumn{4}{c|}{$S$-$D$ wave mixing effect}&\multicolumn{4}{c}{Coupled channel}\\\midrule[1.0pt]
$J^P$&$\Lambda$ &$E$  &$r_{\rm RMS}$ &$\Lambda$ &$E$  &$r_{\rm RMS}$ &P(${}^4\mathbb{S}_{\frac{3}{2}}/{}^2\mathbb{D}_{\frac{3}{2}}/{}^4\mathbb{D}_{\frac{3}{2}})$&$\Lambda$ &$E$  &$r_{\rm RMS}$ &P($\Omega_{c}\bar D_s^*/\Omega_{c}^{*}\bar D_s^*$)\\
\hline
\multirow{3}{*}{${3}/{2}^{-}$}&1.96&$-0.19$ &4.76        &1.96&$-0.33$ &4.14&\textbf{99.94}/0.01/0.05         &1.67&$-1.36$ &2.27&\textbf{95.84}/4.16  \\
                              &1.98&$-5.36$ &1.09        &1.98&$-5.71$ &1.06&\textbf{99.92}/0.02/0.06         &1.69&$-8.38$ &0.90&\textbf{92.17}/7.83     \\
                              &1.99&$-9.44$&0.82       &1.99&$-9.84$ &0.81&\textbf{99.93}/0.02/0.05         &1.70&$-13.35$ &0.72&\textbf{91.00}/9.00     \\
\bottomrule[1.0pt]\bottomrule[1.0pt]
\end{tabular}
\end{table*}

For the $S$-wave $\Omega_{c}\bar D_s^*$ state with $J^P={3}/{2}^{-}$, we notice that the effective potentials from the $f_0$, $\eta$, and $\phi$ exchanges provide the attractive forces, and there exist the bound state solutions with the cutoff parameter around 1.96 GeV by performing the single channel analysis. More interestingly, the bound state properties will change accordingly after including the coupled channels $\Omega_{c}\bar D_s^*$ and $\Omega_{c}^{*}\bar D_s^*$, where we can obtain the loosely bound state solutions when the cutoff parameter $\Lambda$ around 1.67 GeV. Moreover, this bound state is mainly composed of the $\Omega_{c}\bar D_s^*$ channel with the probabilities over 90\%. Based on our numerical results, the $S$-wave $\Omega_{c}\bar D_s^*$ state with $J^P={3}/{2}^{-}$ can be recommended as a good candidate of the hidden-charm molecular pentaquark with triple strangeness.

Comparing the numerical results, it is obvious that the $D$-wave probabilities are less than 1\% and the $S$-$D$ mixing effect can be ignored in forming the $S$-wave $\Omega_{c}\bar D_s^*$ bound states, but the coupled channel effect is obvious in generating the $S$-wave $\Omega_{c}\bar D_s^*$ bound states, especially for the $S$-wave $\Omega_{c}\bar D_s^*$ molecular candidate with $J^P={3}/{2}^{-}$.

For the $S$-wave $\Omega_{c}^{*}\bar D_s^*$ system, the bound state properties are collected in Table~\ref{r4}. Here, we still scan the $\Lambda$ parameter range from 1.00 GeV to 4.00 GeV. For $J^P=1/2^-$, the binding energy is a few MeV and the root-mean-square radii are around 1.00 fm with the cutoff parameter $\Lambda$ larger than 3.59 GeV when only considering the $S$-wave channel, and we can also obtain the bound state solutions when the cutoff value $\Lambda$ is lowered down 3.51 GeV after adding the contribution of the $D$-wave channels. Because the obtained cutoff parameter $\Lambda$ is far away from 1.00 GeV \cite{Tornqvist:1993ng,Tornqvist:1993vu}, our numerical results disfavor the existence of the molecular candidate for the $S$-wave $\Omega_{c}^{*}\bar D_s^*$ state with $J^P=1/2^-$. For $J^P=3/2^-$, there do not exist the bound state solutions when the cutoff parameter varies from 1.00 GeV to 4.00 GeV. This situation does not change when the $S$-$D$ wave mixing effect is considered. Thus, we can exclude the $S$-wave $\Omega_{c}^{*}\bar D_s^*$ state with $J^P=3/2^-$ as the hadronic molecular candidate. For $J^P=5/2^-$, we notice that the total effective potentials due to the $f_0$, $\eta$, and $\phi$ exchanges are attractive. We can obtain the loosely bound state solutions by taking the cutoff value around 1.64 GeV when only considering the contribution of the $S$-wave channel, and the bound state solutions also can be found with the cutoff parameter around 1.64 GeV after considering the $S$-$D$ wave mixing effect. As a result, the $S$-wave $\Omega_{c}^{*}\bar D_s^*$ state with $J^P=5/2^-$ can be regarded as the hidden-charm molecular pentaquark candidate with triple strangeness.
\renewcommand\tabcolsep{0.05cm}
\renewcommand{\arraystretch}{1.50}
\begin{table}[!htbp]
\caption{Bound state solutions of the $S$-wave $\Omega_{c}^{*}\bar D_s^*$ system. The units are the same as Table~\ref{r1}.}\label{r4}
\begin{tabular}{c|ccc|cccc}\toprule[1pt]\toprule[1pt]
\multicolumn{1}{c|}{Effect}&\multicolumn{3}{c|}{Single channel}&\multicolumn{4}{c}{$S$-$D$ wave mixing effect}\\\midrule[1.0pt]
$J^P$&$\Lambda$ &$E$  &$r_{\rm RMS}$ &$\Lambda$ &$E$  &$r_{\rm RMS}$ &P(${}^2\mathbb{S}_{\frac{1}{2}}/{}^4\mathbb{D}_{\frac{1}{2}}/{}^6\mathbb{D}_{\frac{1}{2}})$\\
\cline{1-8}
\multirow{3}{*}{${1}/{2}^{-}$}&$3.59$&$-0.27$&$4.96$&      3.51&$-0.29$ &4.89&\textbf{99.97}/0.02/0.01\\
                              &$3.80$&$-1.18$&$2.96$&      3.76&$-1.74$ &2.52&\textbf{99.92}/0.05/0.03\\
                              &$4.00$&$-2.63$&$2.11$&      4.00&$-4.35$ &1.72&\textbf{99.87}/0.08/0.05\\
\midrule[1.0pt]
$J^P$&$\Lambda$ &$E$  &$r_{\rm RMS}$ &$\Lambda$ &$E$  &$r_{\rm RMS}$ &P(${}^6\mathbb{S}_{\frac{5}{2}}/{}^2\mathbb{D}_{\frac{5}{2}}/{}^4\mathbb{D}_{\frac{5}{2}}/{}^6\mathbb{D}_{\frac{5}{2}})$\\
\cline{1-8}
\multirow{3}{*}{${5}/{2}^{-}$}&1.64&$-0.31$ &4.27&      1.64&$-0.80$ &2.97&\textbf{99.81}/0.02/0.01/0.15\\
                              &1.66&$-4.93$ &1.19&      1.66&$-5.81$ &1.11&\textbf{99.76}/0.03/0.01/0.20\\
                              &1.68&$-13.13$ &0.74&     1.67&$-9.55$ &0.87&\textbf{99.77}/0.03/0.01/0.19\\
\bottomrule[1pt]\bottomrule[1pt]
\end{tabular}
\end{table}

To summarize, we predict two types of hidden-charm molecular pentaquark states with triple strangeness, i.e., the $S$-wave $\Omega_{c}\bar D_s^*$ molecular state with $J^P={3}/{2}^{-}$ and the $S$-wave $\Omega_{c}^{*}\bar D_s^*$ molecular state with $J^P=5/2^-$.
{Here, we want to indicate that the effective potentials from the $\phi$ and $\eta$ exchanges are attractive for the $\Omega_{c}\bar D_s^*$ system with $J^P={3}/{2}^{-}$ and the $\Omega_{c}^{*}\bar D_s^*$ system with $J^P=5/2^-$, which is due to the contributions from the ${\bf q}^2$ terms in the deduced effective potentials. In fact, this issue has been discussed in Ref. \cite{1839195}.}

\section{Decay behaviors of these possible $\Omega_{c}^{(*)}\bar D_s^*$ molecular states}\label{sec3}

In order to further reveal the inner structures and properties of the possible hidden-charm molecular pentaquarks with triple strangeness, we calculate the strong decay behaviors of these possible molecular candidates. In this work, we discuss the hidden-charm decay mode, the corresponding final states including the $\eta_c(1S)\Omega$ and $J/\psi\Omega$. Different with the binding of the possible hidden-charm molecular pentaquarks with triple strangeness, the interactions in the very short range distance contribute to the hidden-charm decay processes. Thus, the quark-interchange model \cite{Barnes:1991em, Barnes:1999hs} can be as a reasonable theoretical framework.

\subsection{The quark-interchange model}

When using the quark-interchange model to estimate the transition amplitudes in calculating the decay widths, we usually adopt the nonrelativistic quark model to describe the quark-quark interaction \cite{Wang:2019spc,Xiao:2019spy}, which is expressed as \cite{Wong:2001td}
\begin{equation}
V_{ij}(q^2)=\frac{\lambda_i}{2}\cdot\frac{\lambda_j}{2}\left(\frac{4\pi\alpha_s}{q^2}+\frac{6\pi b}{q^4}-\frac{8\pi\alpha_s}{3m_im_j}e^{-{\frac{q^2}{4\sigma^2}}}{\bf{s}}_i\cdot{\bf{s}}_j\right),\label{V}
\end{equation}
where $\lambda_i(\lambda_j)$, $m_i(m_j)$, and ${\bf{s}}_i({\bf{s}}_j)$ represent the color factor, the mass, and the spin operator of the interacting quarks, respectively. $\alpha_s$ denotes the running coupling constant, which reads as \cite{Wong:2001td}
\begin{equation}
\alpha_s(Q^2)=\frac{12\pi}{\left(32-2n_f\right){\rm ln}\left(A+\frac{Q^2}{B^2}\right)}, \label{alpha}
\end{equation}
where $Q^2$ is the square of the invariant mass of the interacting quarks, and the relevant parameters \cite{Wong:2001td} in Eqs.~(\ref{V}) and ~(\ref{alpha}) are collected in Table~\ref{parameters}.
\renewcommand\tabcolsep{0.16cm}
\renewcommand{\arraystretch}{1.50}
\begin{table}[!htbp]
\caption{The parameters of the nonrelativistic quark model \cite{Wong:2001td} and the oscillating parameters of the Gaussian function \cite{Wang:2019spc}.}\label{parameters}
\centering
\begin{tabular}{c|ccc}
\toprule[1.0pt]
\toprule[1.0pt]
\multirow{4}{*}{Quark model}&$b~(\rm{GeV}^2)$&$\sigma~(\rm{GeV})$&$A$\\
&0.180&0.897&10\\
\cline{2-4}
&$B$~(GeV)&$m_s~(\rm{GeV})$&$m_c~(\rm{GeV})$\\
&0.310&0.575&1.776  \\
\cline{1-4}\midrule[1pt]
\multirow{6}{*}{Oscillating parameters}&$\beta_{D^{\ast}_s}~(\rm{GeV})$&$\beta_{\eta_{c}}~(\rm{GeV})$&$\beta_{J/\psi}~(\rm{GeV})$\\
 &0.562&0.838&0.729\\
\cline{2-4}
&$\alpha_{\lambda\Omega}~(\rm{GeV})$&$\alpha_{\rho\Omega}~(\rm{GeV})$&$\alpha_{\lambda\Omega_c}~(\rm{GeV})$\\
&0.466&0.407&0.583\\
\cline{2-4}
&$\alpha_{\rho\Omega_c}~(\rm{GeV})$&$\alpha_{\lambda\Omega^{\ast}_c}~(\rm{GeV})$&$\alpha_{\rho\Omega^{\ast}_c}~(\rm{GeV})$\\
&0.444&0.537&0.423\\
\bottomrule[1.0pt]
\bottomrule[1.0pt]
\end{tabular}
\end{table}

\begin{figure}[!htbp]
\centering
\begin{tabular}{c}
\includegraphics[width=8.4cm,keepaspectratio]{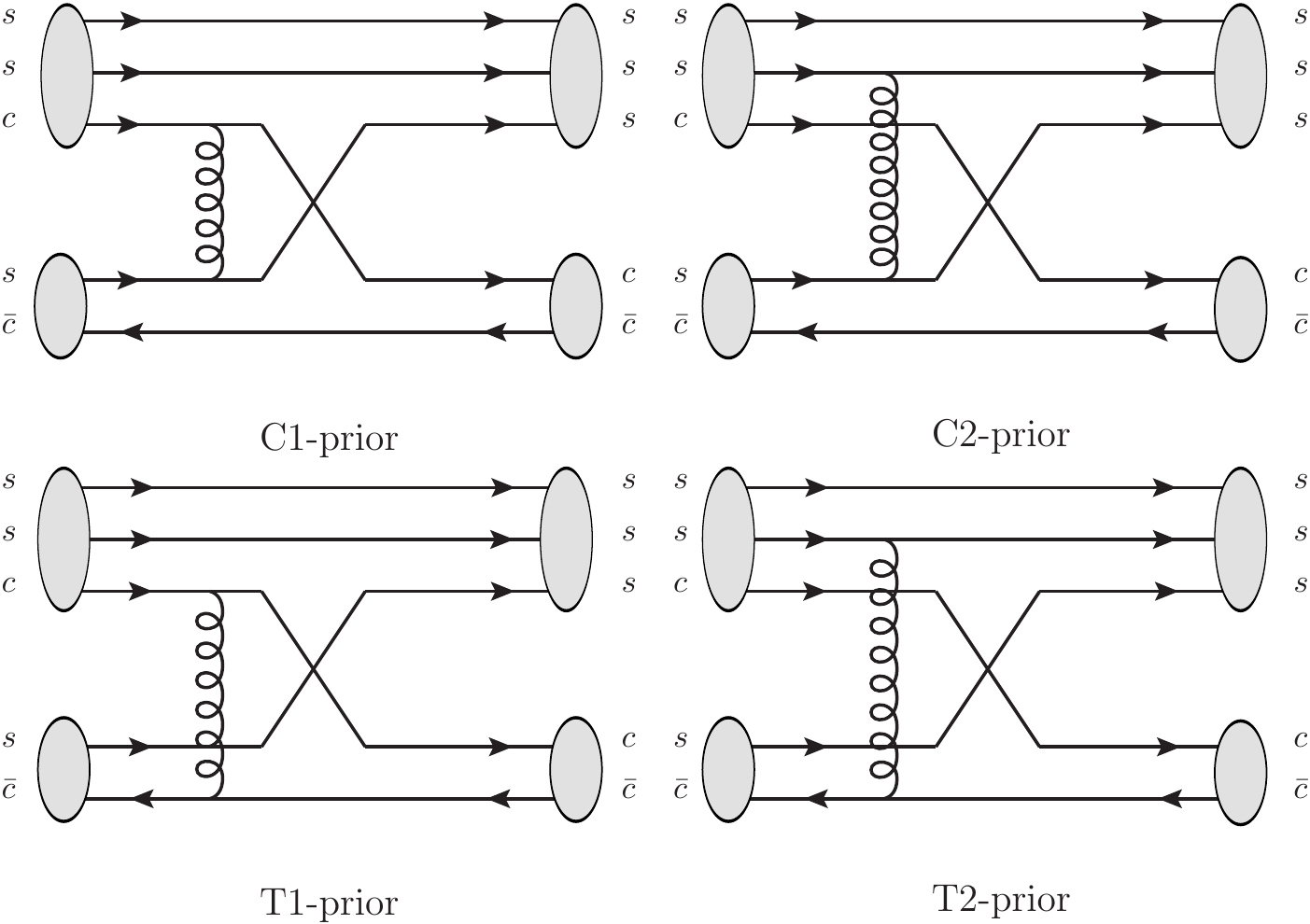}
\end{tabular}
\caption{Quark-interchange diagrams for the process $A(css)+B(s\bar{c})\to C(sss)+D(c\bar{c})$ in the hadronic molecular picture.}\label{interchangediagrams}
\end{figure}

To get the transition amplitudes within the quark-interchange model, we take the same convention as the previous work \cite{Wang:2019spc,Xiao:2019spy,Wang:2020prk}. The transition amplitude for the process $A(css)+B(s\bar{c})\to C(sss)+D(c\bar{c})$ can be decomposed as four processes in the hadronic molecular picture, which are illustrated in Fig.~\ref{interchangediagrams}.

The Hamiltonian of the initial hidden-charm molecular pentaquark state can be written as \cite{Wang:2019spc}
\begin{equation}
H_{\rm {Initial}}=H^0_A+H^0_B+V_{AB},
\end{equation}
where $H^0_A$ and $H^0_B$ are the Hamiltonian of the free baryon A and meson B, and $V_{AB}$ denotes the interaction between the baryon A and the meson B.

Furthermore, we define the color wave function $\omega_{\rm{color}}$, the flavor wave function $\chi_{\rm{flavor}}$, the spin wave function $\chi_{\rm{spin}}$, and the momentum space wave function $\phi(\bf{p})$, respectively. Thus, the total wave function can be expressed as
\begin{equation}
\psi_{\rm{total}}=\omega_{\rm{color}}\chi_{\rm{flavor}}\chi_{\rm{spin}}\phi(\bf{p}).\label{wavefunction}
\end{equation}
In this work, we take the Gaussian functions to approximate the momentum space wave functions for the baryon, meson, and molecule. The more explicit forms of the relevant Gaussian function can be found in Ref. \cite{Wang:2019spc}, and the oscillating parameters of the meson and baryon are estimated by fitting their mass spectrum in the Godfrey-Isgur model \cite{Godfrey:1985xj}, which are listed in Table~\ref{parameters}. For an $S$-wave loosely bound state composed of two hadrons A and B, the oscillating parameter $\beta$ can be related to the mass of the molecular state $m$, i.e., $\beta=\sqrt{3 \mu (m_A+m_B-m)}$ with $\mu=\frac{m_Am_B}{m_A+m_B}$ \cite{Weinberg:1962hj,Weinberg:1963zza,Guo:2017jvc}.
And then, the $T$-matrix $T_{fi}$ represents the relevant effective potential in the quark-interchange diagrams, which can be factorized as
\begin{equation}
T_{fi}=I_{\rm{color}}I_{\rm{flavor}}I_{\rm{spin}}I_{\rm{space}},
\end{equation}
where $I_{i}$ with the subscripts color, flavor, spin, and space stand for the corresponding factors, and the calculation details of these factors $I_{i} \,(i=\rm{color}, \,\rm{flavor}, \,\rm{spin}, \,\rm{space})$ are referred to in Ref.~\cite{Wang:2019spc}.

For the two-body strong decay widths of these discussed molecular candidates, they can be explicitly expressed as
\begin{eqnarray}
\Gamma=\frac{|{\bf{P}}_C|}{32\pi^2m^2(2J+1)}\int d\Omega|\mathcal{M}|^2.\label{width}
\end{eqnarray}
In the above expression, ${\bf{P}}_C$, $m$, and $\mathcal{M}$ stand for the momentum of the final state, the mass of the molecular state, and the transition amplitude of the discussed process, respectively. Here, we want to emphasize that there exists a relation of the transition amplitude $\mathcal{M}$ and the $T$-matrix $T_{fi}$, i.e.,
\begin{eqnarray}
\mathcal{M}=-(2\pi)^{\frac{3}{2}}\sqrt{2m2E_C2E_D}T_{fi},\label{M}
\end{eqnarray}
where $E_C$ and $E_D$ are the energies of the final states C and D, respectively. Through the above preparation, we can calculate the two-body hidden-charm strong decay widths of these proposed $\Omega_{c}^{(*)}\bar D_s^*$ molecular states.

\subsection{Two-body hidden-charm strong decay widths of these proposed $\Omega_{c}^{(*)}\bar D_s^*$ molecular states}

In the above section, our results suggest that the $S$-wave $\Omega_{c}\bar D_s^*$ state with $J^P={3}/{2}^{-}$ and the $S$-wave $\Omega_{c}^{*}\bar D_s^*$ state with $J^P=5/2^-$ can be regarded as the hidden-charm molecular pentaquark candidates with triple strangeness. Thus, we will study the two-body strong decay property of these possible hidden-charm molecular pentaquarks with triple strangeness, which provides valuable information to search for these proposed molecular candidates in experiment.

In this work, we focus on the two-body hidden-charm strong decay channels for these predicted hidden-charm molecular pentaquarks with triple strangeness. For the $S$-wave $\Omega_{c}\bar D_s^*$ molecular state with $J^P={3}/{2}^{-}$, it can decay into the $J/\psi \, \Omega$ and $\eta_c \,\Omega$ channels through the $S$-wave interaction. For the $S$-wave $\Omega_{c}^{*}\bar D_s^*$ molecular state with $J^P={5}/{2}^{-}$, we only take into account the $J/\psi \,\Omega$ decay channel via the $S$-wave coupling, while the $\eta_c \,\Omega$ channel is suppressed since it is a $D$-wave decay \cite{Wang:2019spc}.

\begin{figure}[!htbp]
\includegraphics[width=4.1cm,keepaspectratio]{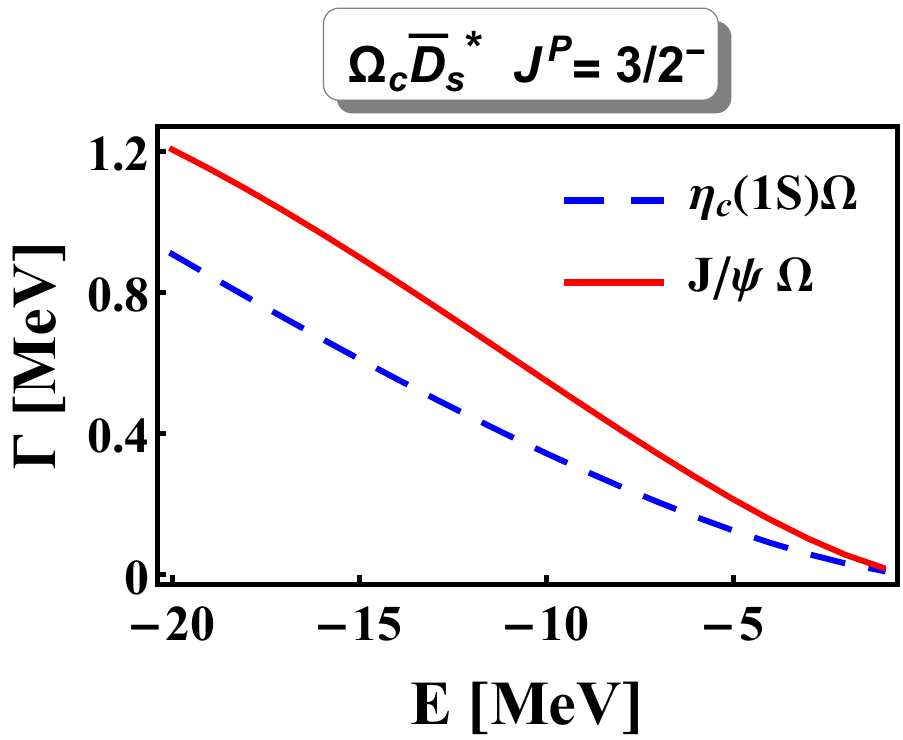}~
\includegraphics[width=4.1cm,keepaspectratio]{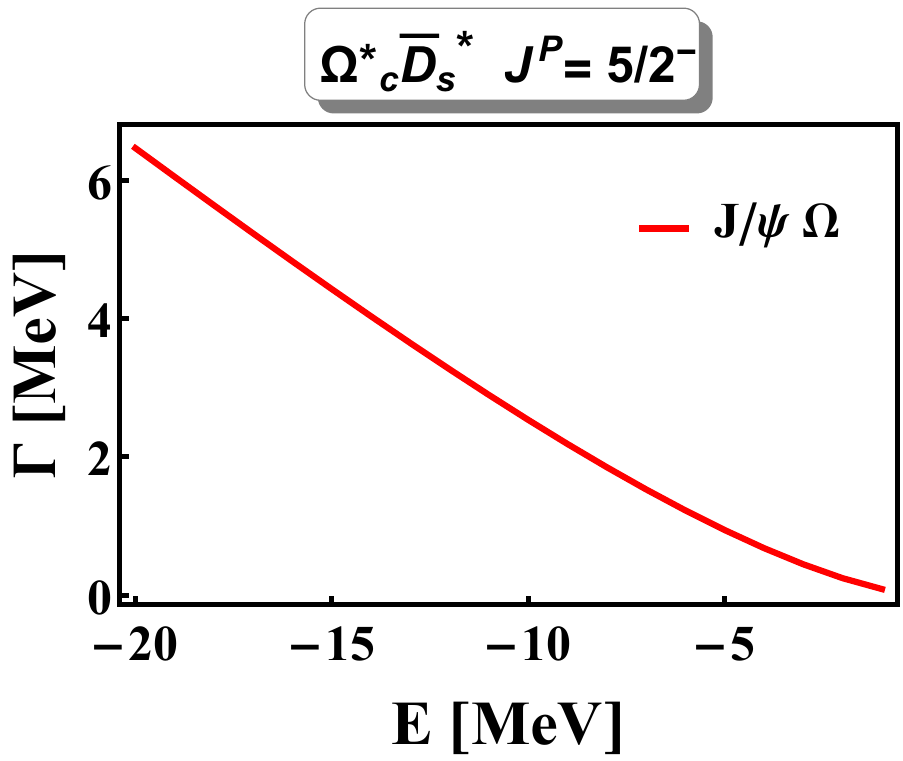}
\caption{The binding energies dependence of the decay widths for the $S$-wave $\Omega_{c}\bar D_s^*$ molecular state with $J^P={3}/{2}^{-}$ and the $S$-wave $\Omega_{c}^{*}\bar D_s^*$ molecular state with $J^P={5}/{2}^{-}$.}
\label{decay}
\end{figure}

In order to intuitively clarify the uncertainty of the binding energies, we present the binding energies dependence of the decay widths for the $S$-wave $\Omega_{c}\bar D_s^*$ molecular state with $J^P={3}/{2}^{-}$ and the $S$-wave $\Omega_{c}^{*}\bar D_s^*$ molecular state with $J^P={5}/{2}^{-}$ in Fig. \ref{decay}. As stressed in Sec. \ref{sec2}, the hadronic molecule is a loosely bound state \cite{Chen:2017xat}, so the binding energies of these hidden-charm molecular pentaquarks with triple strangeness change from $-20$ to $-1$ MeV in calculating the decay widths. With increasing the absolute values of the binding energy, the decay widths become larger, which is consistent with other theoretical calculations \cite{Chen:2017xat,Lin:2017mtz,Lin:2018kcc,Lin:2018nqd,Shen:2019evi,Lin:2019qiv,Lin:2019tex,Dong:2019ofp,Dong:2020rgs,Xiao:2019mvs,Wu:2018xaa,Chen:2017abq,Xiao:2016mho}.

As illustrated in Fig. \ref{decay}, when the binding energies are taken as $-15$ MeV with typical values, the dominant decay channel is the $J/\psi \, \Omega$ around one MeV for the $S$-wave $\Omega_{c}\bar D_s^*$ molecular state with $J^P={3}/{2}^{-}$, and the decay width of the $J/\psi \, \Omega$ channel is predicted to be around several MeV for the $S$-wave $\Omega_{c}^{*}\bar D_s^*$ molecule with $J^P={5}/{2}^{-}$. Thus, the $J/\psi \,\Omega$ should be the promising channel to observe the $S$-wave $\Omega_{c}\bar D_s^*$ molecular state with $J^P={3}/{2}^{-}$ and the $S$-wave $\Omega_{c}^{*}\bar D_s^*$ molecular state with $J^P={5}/{2}^{-}$. Meanwhile, it is interesting to note that the $S$-wave $\Omega_{c}\bar D_s^*$ molecular state with $J^P={3}/{2}^{-}$ prefers to decay into the $J/\psi \,\Omega$ channel, but the decay width of the $\eta_c\,\Omega$ channel is comparable to the $J/\psi \,\Omega$ channel, which indicates that the $S$-wave $\Omega_{c}\bar D_s^*$ molecule with $J^P={3}/{2}^{-}$ can be detected in the $\eta_c \,\Omega$ channel in future experiment.

{{In the heavy quark symmetry, the relative partial decay branch ratio between the $\eta_c(1S)\Omega$ and $J/\psi\Omega$ for the $\Omega_c\bar{D}_s^*$ state with $J^P=3/2^-$ can be estimated as
\begin{eqnarray}
\mathcal{R}_{\text{HQS}}=\frac{\Gamma(\Omega_c\bar{D}_s^*\to\eta_c(1S)\Omega)}{\Gamma(\Omega_c\bar{D}_s^*\to J/\psi\Omega)}=0.6,
\end{eqnarray}
since the relative momentum in the $\eta_c(1S)\Omega$ channel is larger than that in the $J/\psi\Omega$ channel, $\mathcal{R}(E)$ should be a little larger than $\mathcal{R}_{\text{HQS}}=0.6$, where $E$ is the binding energy. In our calculation, we obtain
\begin{eqnarray*}
\mathcal{R}(-5~\text{MeV})=\mathcal{R}(-10~\text{MeV})=0.62, \quad \mathcal{R}(-15~\text{MeV})=0.67.
\end{eqnarray*}
Obviously, our results are consistent with the estimation in the heavy quark limit.}}

\section{Summary}\label{sec4}
Searching for exotic hadronic state is an interesting and important research topic of hadron physics. With accumulation of experimental data, the LHCb observed three narrow $P_c(4312)$, $P_c(4440)$, and $P_c(4457)$ in 2019 \cite{Aaij:2019vzc}, and found the evidence of the $P_{cs}(4459)$ as a hidden-charm pentaquark with strangeness \cite{Aaij:2020gdg}. These progresses make us have reason to believe that there should exist a zoo of the hidden-charm molecular pentaquark. At present, the hidden-charm molecular pentaquark with triple strangeness is still missing, which inspires our interest in exploring how to find these intriguing hidden-charm molecular pentaquark states with triple strangeness.

Mass spectrum information is crucial to searching for them. In this work, we perform the dynamical calculation of the possible hidden-charm molecular pentaquark states with triple strangeness from the $\Omega_{c}^{(*)}\bar{D}_s^{(*)}$ interactions, where the effective potentials can be obtained by the OBE model. By finding bound state solutions of these discussed systems, we find that the most promising hidden-charm molecular pentaquarks with triple strangeness are the $S$-wave $\Omega_{c}\bar D_s^*$ state with $J^P={3}/{2}^{-}$ and the $S$-wave $\Omega_{c}^{*}\bar D_s^*$ state with $J^P=5/2^-$. Besides mass spectrum study, we also discuss their two-body hidden-charm strong decay behaviors within the quark-interchange model. In concrete calculation, we mainly focus on the $J/\psi \,\Omega$ and $\eta_c \,\Omega$ decay modes for the predicted $S$-wave $\Omega_{c}\bar D_s^*$ molecule with $J^P={3}/{2}^{-}$ and the $J/\psi \,\Omega$ decay channel for the predicted $S$-wave $\Omega_{c}^{*}\bar D_s^*$ molecule with $J^P={5}/{2}^{-}$.

In the following years, the LHCb Collaboration will collect more experimental data at Run III and upgrade the High-Luminosity-LHC \cite{Bediaga:2018lhg}. Experimental searches for these predicted hidden-charm molecular pentaquarks with triple strangeness are an area full of opportunities and challenges in future experiments.

\section*{ACKNOWLEDGMENTS}
We would like to thank Z. W. Liu, G. J. Wang, L. Y. Xiao, and S. Q. Luo for very helpful discussions. This work is supported by the China National Funds for Distinguished Young Scientists under Grant No. 11825503, National Key Research and Development Program of China under Contract No. 2020YFA0406400, the 111 Project under Grant No. B20063, and the National Natural Science Foundation of China under Grant No. 12047501.
R. C. is supported by the National Postdoctoral Program for Innovative Talent.

\appendix
\section{Relevant subpotentials}\label{app01}
Through the standard strategy \cite{Wang:2020dya,Wang:2019nwt,Wang:2019aoc}, we can derive the effective potentials in the coordinate space for these investigated $\Omega_{c}^{(*)}\bar{D}_s^{(*)}$ systems, i.e.,
\begin{eqnarray}
\mathcal{V}^{\Omega_{c}\bar D_s\rightarrow\Omega_{c}\bar D_s}&=&-AY_{f_0}-\frac{C}{2}Y_{\phi},\\
\mathcal{V}^{\Omega_{c}^*\bar D_s\rightarrow\Omega_{c}^*\bar D_s}&=&-A\mathcal{A}_{1}Y_{f_0}-\frac{C}{2}\mathcal{A}_{1}Y_{\phi},\\
\mathcal{V}^{\Omega_{c}\bar D_s^*\rightarrow\Omega_{c}\bar D_s^*}&=&-A\mathcal{A}_{2}Y_{f_0}+\frac{2B}{9}\left[\mathcal{A}_{3}\mathcal{O}_r+\mathcal{A}_{4}\mathcal{P}_r\right]Y_{\eta}\nonumber\\
&&-\frac{C}{2}\mathcal{A}_{2}Y_{\phi}-\frac{2D}{9}\left[2\mathcal{A}_{3}\mathcal{O}_r-\mathcal{A}_{4}\mathcal{P}_r\right]Y_{\phi},\\
\mathcal{V}^{\Omega_{c}^*\bar D_s^*\rightarrow\Omega_{c}^{*}\bar D_s^*}&=&-A\mathcal{A}_{5}Y_{f_0}-\frac{B}{3}\left[\mathcal{A}_{6}\mathcal{O}_r+\mathcal{A}_{7}\mathcal{P}_r\right]Y_{\eta}\nonumber\\
&&-\frac{C}{2}\mathcal{A}_{5}Y_{\phi}+\frac{D}{3}\left[2\mathcal{A}_{6}\mathcal{O}_r-\mathcal{A}_{7}\mathcal{P}_r\right]Y_{\phi},\\
\mathcal{V}^{\Omega_{c}\bar D_s\rightarrow\Omega_{c}^*\bar D_s}&=&\frac{A}{\sqrt{3}}\mathcal{A}_8Y_{f_01}+\frac{C}{2\sqrt{3}}\mathcal{A}_8Y_{\phi1},\\
\mathcal{V}^{\Omega_{c}\bar D_s\rightarrow\Omega_{c}\bar D_s^*}&=&\frac{2B}{9}\left[\mathcal{A}_9\mathcal{O}_r+\mathcal{A}_{10}\mathcal{P}_r\right]Y_{\eta2}\nonumber\\
&&+\frac{2D}{9}\left[2\mathcal{A}_9\mathcal{O}_r-\mathcal{A}_{10}\mathcal{P}_r\right]Y_{\phi2},\\
\mathcal{V}^{\Omega_{c}\bar D_s\rightarrow\Omega_{c}^*\bar D_s^*}&=&-\frac{B}{3\sqrt{3}}\left[\mathcal{A}_{11}\mathcal{O}_r+\mathcal{A}_{12}\mathcal{P}_r\right]Y_{\eta3}\nonumber\\
&&-\frac{D}{3\sqrt{3}}\left[2\mathcal{A}_{11}\mathcal{O}_r-\mathcal{A}_{12}\mathcal{P}_r\right]Y_{\phi3},\\
\mathcal{V}^{\Omega_{c}^*\bar D_s\rightarrow\Omega_{c}\bar D_s^*}&=&\frac{B}{3\sqrt{3}}\left[\mathcal{A}_{13}\mathcal{O}_r+\mathcal{A}_{14}\mathcal{P}_r\right]Y_{\eta4}\nonumber\\
&&+\frac{D}{3\sqrt{3}}\left[2\mathcal{A}_{13}\mathcal{O}_r-\mathcal{A}_{14}\mathcal{P}_r\right]Y_{\phi4},\\
\mathcal{V}^{\Omega_{c}^*\bar D_s\rightarrow\Omega_{c}^*\bar D_s^*}&=&\frac{B}{3}\left[\mathcal{A}_{15}\mathcal{O}_r+\mathcal{A}_{16}\mathcal{P}_r\right]Y_{\eta5}\nonumber\\
&&+\frac{D}{3}\left[2\mathcal{A}_{15}\mathcal{O}_r-\mathcal{A}_{16}\mathcal{P}_r\right]Y_{\phi5},\\
\mathcal{V}^{\Omega_{c}\bar D_s^*\rightarrow\Omega_{c}^*\bar D_s^*}&=&\frac{A}{\sqrt{3}}\mathcal{A}_{17}Y_{f_06}+\frac{B}{3\sqrt{3}}\left[\mathcal{A}_{18}\mathcal{O}_r+\mathcal{A}_{19}\mathcal{P}_r\right]Y_{\eta6}\nonumber\\
&&+\frac{C\mathcal{A}_{17}}{2\sqrt{3}}Y_{\phi6}-\frac{D}{3\sqrt{3}}\left[2\mathcal{A}_{18}\mathcal{O}_r-\mathcal{A}_{19}\mathcal{P}_r\right]Y_{\phi6}.\nonumber\\
\end{eqnarray}
Here, $\mathcal{O}_r = \frac{1}{r^2}\frac{\partial}{\partial r}r^2\frac{\partial}{\partial r}$ and $\mathcal{P}_r = r\frac{\partial}{\partial r}\frac{1}{r}\frac{\partial}{\partial r}$. Additionally, we also define several variables, which include $A=l_Sg_S$, $B=g_1 g/f_\pi^2$, $C=\beta_S \beta g_{V}^2$, and $D=\lambda_S \lambda g_V^2$. The function $Y_i$ can be defined as
\begin{eqnarray}
Y_i= \dfrac{e^{-m_ir}-e^{-\Lambda_ir}}{4\pi r}-\dfrac{\Lambda_i^2-m_i^2}{8\pi\Lambda_i}e^{-\Lambda_ir}.
\end{eqnarray}
Here, $m_i=\sqrt{m^2-q_i^2}$ and $\Lambda_i=\sqrt{\Lambda^2-q_i^2}$. Variables $q_i\,(i = 1\,, . . . ,\, 6)$ are defined as $q_1=0.04$ GeV, $q_2=0.06$ GeV, $q_3=0.02$ GeV, $q_4=0.10$ GeV, $q_5=0.06$ GeV, and $q_6=0.04$ GeV.

In the above effective potentials, we also introduce several operators, i.e.,
\begin{eqnarray}
\mathcal{A}_{1}&=&\sum_{a,b,m,n}C^{\frac{3}{2},a+b}_{\frac{1}{2}a,1b}C^{\frac{3}{2},m+n}_{\frac{1}{2}m,1n}\chi^{\dagger a}_{3}\left({\bm\epsilon^{\dagger b}_{3}}\cdot{\bm\epsilon^{n}_{1}}\right)\chi^{m}_1,\nonumber\\
\mathcal{A}_{2}&=&\chi^{\dagger}_3\left({\bm\epsilon^{\dagger}_{4}}\cdot{\bm\epsilon_{2}}\right)\chi_1,\nonumber\\
\mathcal{A}_{3}&=&\chi^{\dagger}_3\left[{\bm\sigma}\cdot\left(i{\bm\epsilon_{2}}\times{\bm\epsilon^{\dagger}_{4}}\right)\right]\chi_1,\nonumber\\
\mathcal{A}_{4}&=&\chi^{\dagger}_3T({\bm\sigma},i{\bm\epsilon_{2}}\times{\bm\epsilon^{\dagger}_{4}})\chi_1,\nonumber\\
\mathcal{A}_{5}&=&\sum_{a,b,m,n}C^{\frac{3}{2},a+b}_{\frac{1}{2}a,1b}C^{\frac{3}{2},m+n}_{\frac{1}{2}m,1n}\chi^{\dagger a}_3\left({\bm\epsilon^{n}_{1}}\cdot{\bm\epsilon^{\dagger b}_{3}}\right)\left({\bm\epsilon_{2}}\cdot{\bm\epsilon^{\dagger}_{4}}\right)\chi^m_1,\nonumber\\
\mathcal{A}_{6}&=&\sum_{a,b,m,n}C^{\frac{3}{2},a+b}_{\frac{1}{2}a,1b}C^{\frac{3}{2},m+n}_{\frac{1}{2}m,1n}\chi^{\dagger a}_3\left({\bm\epsilon^{n}_{1}}\times{\bm\epsilon^{\dagger b}_{3}}\right)\cdot\left({\bm\epsilon_{2}}\times{\bm\epsilon^{\dagger}_{4}}\right)\chi^m_1,\nonumber\\
\mathcal{A}_{7}&=&\sum_{a,b,m,n}C^{\frac{3}{2},a+b}_{\frac{1}{2}a,1b}C^{\frac{3}{2},m+n}_{\frac{1}{2}m,1n}\chi^{\dagger a}_3T({\bm\epsilon^{n}_{1}}\times{\bm\epsilon^{\dagger b}_{3}},{\bm\epsilon_{2}}\times{\bm\epsilon^{\dagger}_{4}})\chi^m_1,\nonumber\\
\mathcal{A}_{8}&=&\sum_{a,b}C^{\frac{3}{2},a+b}_{\frac{1}{2}a,1b}\chi^{\dagger a}_{3}\left({\bm\epsilon^{\dagger b}_{3}}\cdot{\bm\sigma}\right)\chi_1,\nonumber\\
\mathcal{A}_{9}&=&\chi^{\dagger}_3\left({\bm\sigma}\cdot{\bm\epsilon^{\dagger}_{4}}\right)\chi_1,\nonumber\\
\mathcal{A}_{10}&=&\chi^{\dagger}_3T({\bm\sigma},{\bm\epsilon^{\dagger}_{4}})\chi_1,\nonumber\\
\mathcal{A}_{11}&=&\sum_{a,b}C^{\frac{3}{2},a+b}_{\frac{1}{2}a,1b}\chi^{\dagger a}_3\left[{\bm\epsilon^{\dagger}_{4}}\cdot\left(i{\bm\sigma}\times{\bm\epsilon^{\dagger b}_{3}}\right)\right]\chi_1,\nonumber\\
\mathcal{A}_{12}&=&\sum_{a,b}C^{\frac{3}{2},a+b}_{\frac{1}{2}a,1b}\chi^{\dagger a}_3T({\bm\epsilon^{\dagger}_{4}},i{\bm\sigma}\times{\bm\epsilon^{\dagger b}_{3}})\chi_1,\nonumber\\
\mathcal{A}_{13}&=&\sum_{a,b}C^{\frac{3}{2},a+b}_{\frac{1}{2}a,1b}\chi^{\dagger}_3\left[{\bm\epsilon^{\dagger}_{4}}\cdot\left(i{\bm\sigma}\times{\bm\epsilon^{b}_{1}}\right)\right]\chi^a_1,\nonumber\\
\mathcal{A}_{14}&=&\sum_{a,b}C^{\frac{3}{2},a+b}_{\frac{1}{2}a,1b}\chi^{\dagger}_3T({\bm\epsilon^{\dagger}_{4}},i{\bm\sigma}\times{\bm\epsilon^{b}_{1}})\chi^a_1,\nonumber\\
\mathcal{A}_{15}&=&\sum_{a,b,m,n}C^{\frac{3}{2},a+b}_{\frac{1}{2}a,1b}C^{\frac{3}{2},m+n}_{\frac{1}{2}m,1n}\chi^{\dagger a}_3\left[{\bm\epsilon^{\dagger}_{4}}\cdot\left(i{\bm\epsilon^n_{1}}\times{\bm\epsilon^{\dagger b}_{3}}\right)\right]\chi^m_1,\nonumber\\
\mathcal{A}_{16}&=&\sum_{a,b,m,n}C^{\frac{3}{2},a+b}_{\frac{1}{2}a,1b}C^{\frac{3}{2},m+n}_{\frac{1}{2}m,1n}\chi^{\dagger a}_3T({\bm\epsilon^{\dagger}_{4}},i{\bm\epsilon^n_{1}}\times{\bm\epsilon^{\dagger b}_{3}})\chi^m_1,\nonumber\\
\mathcal{A}_{17}&=&\sum_{a,b}C^{\frac{3}{2},a+b}_{\frac{1}{2}a,1b}\chi^{\dagger a}_3\left({\bm\sigma}\cdot{\bm\epsilon^{\dagger b}_{3}}\right)\left({\bm\epsilon_{2}}\cdot{\bm\epsilon^{\dagger}_{4}}\right)\chi_1,\nonumber\\
\mathcal{A}_{18}&=&\sum_{a,b}C^{\frac{3}{2},a+b}_{\frac{1}{2}a,1b}\chi^{\dagger a}_3\left({\bm\sigma}\times{\bm\epsilon^{\dagger b}_{3}}\right)\cdot\left({\bm\epsilon_{2}}\times{\bm\epsilon^{\dagger}_{4}}\right)\chi_1,\nonumber\\
\mathcal{A}_{19}&=&\sum_{a,b}C^{\frac{3}{2},a+b}_{\frac{1}{2}a,1b}\chi^{\dagger a}_3T({\bm\sigma}\times{\bm\epsilon^{\dagger b}_{3}},{\bm\epsilon_{2}}\times{\bm\epsilon^{\dagger}_{4}})\chi_1.
\end{eqnarray}
Here, $T({\bm x},{\bm y})= 3\left(\hat{\bm r} \cdot {\bm x}\right)\left(\hat{\bm r} \cdot {\bm y}\right)-{\bm x} \cdot {\bm y}$ is the tensor force operator. In Table~\ref{matrix}, we collect the numerical matrix elements $\langle f|\mathcal{A}_k|i\rangle\,(k = 1\,, . . . ,\, 7)$ with the $S$-$D$ wave mixing effect analysis. Of course, the relevant numerical matrix elements $\langle f|\mathcal{A}_k|i\rangle\,(k = 8\,, . . . ,\, 19)$ will be involved in the coupled channel analysis. For the coupled channel analysis with $J=1/2$, we have $\mathcal{A}_{9}=\sqrt{3}$, $\mathcal{A}_{11}=\sqrt{2}$, $\mathcal{A}_{18}=-\sqrt{{2}/{3}}$, and $\mathcal{A}_{k}=0\,(k = 10, \,12, \,17, \,19)$. And, there exists $\mathcal{A}_{13}=1$, $\mathcal{A}_{15}=\sqrt{{5}/{3}}$, $\mathcal{A}_{18}=-\sqrt{{5}/{3}}$, and $\mathcal{A}_{k}=0\,(k = 14, \,16, \,17, \,19)$ for the coupled channel analysis with $J=3/2$.
\renewcommand\tabcolsep{0.60cm}
\renewcommand{\arraystretch}{1.50}
\begin{table*}[htbp]
  \caption{The numerical matrix elements $\langle f|\mathcal{A}_k|i\rangle\,(k = 1\,, . . . ,\, 7)$ with the $S$-$D$ wave mixing effect analysis.}\label{matrix}
  \begin{tabular}{c|ccc}\toprule[1pt]\toprule[1pt]
   {{{Matrix elements}}} & $J=1/2$     & $J=3/2$     & $J=5/2$        \\\midrule[1pt]
 $\langle\Omega_{c}^*\bar D_s|\mathcal{A}_{1}|\Omega_{c}^*\bar D_s\rangle$
            &$/$ &diag(1,1) &$/$                                  \\
 $\langle\Omega_{c}\bar D_s^*|\mathcal{A}_{2}|\Omega_{c}\bar D_s^*\rangle$
            &diag(1,1) &diag(1,1,1) &$/$                        \\

 $\langle\Omega_{c}\bar D_s^*|\mathcal{A}_{3}|\Omega_{c}\bar D_s^*\rangle$
            &diag($-2$,$1$) &diag($1$,$-2$,$1$) &$/$            \\

  $\langle\Omega_{c}\bar D_s^*|\mathcal{A}_{4}|\Omega_{c}\bar D_s^*\rangle$
           &$\left(\begin{array}{cc} 0 & -\sqrt{2} \\ -\sqrt{2} & -2\end{array}\right)$&$\left(\begin{array}{ccc} 0 & 1& 2 \\ 1 & 0& -1 \\ 2 & -1& 0 \end{array}\right)$&$/$           \\

 $\langle\Omega_{c}^{*}\bar D_s^*|\mathcal{A}_{5}|\Omega_{c}^{*}\bar D_s^*\rangle$
            &diag(1,1,1) &diag(1,1,1,1) &diag(1,1,1,1)  \\

 $\langle\Omega_{c}^{*}\bar D_s^*|\mathcal{A}_{6}|\Omega_{c}^{*}\bar D_s^*\rangle$
            &diag($\frac{5}{3}$,$\frac{2}{3}$,$-1$) &diag($\frac{2}{3}$,$\frac{5}{3}$,$\frac{2}{3}$,$-1$) &diag($-1$,$\frac{5}{3}$,$\frac{2}{3}$,$-1$)  \\

 $\langle\Omega_{c}^{*}\bar D_s^*|\mathcal{A}_{7}|\Omega_{c}^{*}\bar D_s^*\rangle$
           &$\left(\begin{array}{ccc} 0 & -\frac{7}{3\sqrt{5}}& \frac{2}{\sqrt{5}} \\ -\frac{7}{3\sqrt{5}} & \frac{16}{15}& -\frac{1}{5} \\ \frac{2}{\sqrt{5}} &-\frac{1}{5}& \frac{8}{5} \end{array}\right)$
           &$\left(\begin{array}{cccc} 0 & \frac{7}{3\sqrt{10}}& -\frac{16}{15}& -\frac{\sqrt{7}}{5\sqrt{2}}\\ \frac{7}{3\sqrt{10}} & 0& -\frac{7}{3\sqrt{10}} & -\frac{2}{\sqrt{35}} \\ -\frac{16}{15} & -\frac{7}{3\sqrt{10}}& 0& -\frac{1}{\sqrt{14}} \\-\frac{\sqrt{7}}{5\sqrt{2}}&-\frac{2}{\sqrt{35}} &-\frac{1}{\sqrt{14}}&\frac{4}{7}\end{array}\right)$
           &$\left(\begin{array}{cccc} 0 & \frac{2}{\sqrt{15}}& \frac{\sqrt{7}}{5\sqrt{3}}& -\frac{2\sqrt{14}}{5}\\ \frac{2}{\sqrt{15}} & 0& \frac{\sqrt{7}}{3\sqrt{5}} & -\frac{4\sqrt{2}}{\sqrt{105}} \\ \frac{\sqrt{7}}{5\sqrt{3}} & \frac{\sqrt{7}}{3\sqrt{5}}& -\frac{16}{21}& -\frac{\sqrt{2}}{7\sqrt{3}} \\-\frac{2\sqrt{14}}{5}&-\frac{4\sqrt{2}}{\sqrt{105}} &-\frac{\sqrt{2}}{7\sqrt{3}}&-\frac{4}{7}\end{array}\right)$\\
           \bottomrule[1pt]\bottomrule[1pt]
  \end{tabular}
\end{table*}

\end{document}